# Predicting Short Response Ratings with Non-Content Related Features: A Hierarchical Modeling Approach

Aubrey Condor, University of California, Berkeley, aubrey_condor@berkeley.edu

**Abstract:** We explore whether the human ratings of open-ended responses can be explained with non-content related features, and if such effects vary across different mathematics-related items. When scoring is rigorously defined and rooted in a measurement framework, educators intend that the features of a response which are indicative of the respondent's level of ability are contributing to scores. However, we find that features such as response length, a grammar score of the response, and a metric relating to key phrase frequency are significant predictors for response ratings. Although our findings are not causally conclusive, they may propel us to be more critical of the way in which we assess open-ended responses, especially in high-stakes scenarios. Educators take great care to provide unbiased, consistent ratings, but it may be that extraneous features unrelated to those which were intended to be rated are being evaluated.

## Introduction

Educational assessment is a key part of instruction and learning, with various types of items allowing teachers to assess student understanding. Open-ended items may be especially beneficial to student learning due to the generation effect (Bertsch et al., 2007) or in combination with self-explanation (Chi et al., 1994) and have specifically been shown to enhance mathematics learning (Hancock, 1995). However, ratings produced by humans often vary in consistency and reliability (Wind & Peterson, 2018). Further, grading open-ended items is time consuming (Hancock, 1995) and while educators are often dealing with limited time-resources, the quality of scoring may be of concern. In this paper, we explore if expert ratings of student responses to mathematics questions can be predicted with non-content related features. In the case where scoring is rooted in a measurement framework (Wilson, 2023) including rigorously defined rubrics with content-related instructions for evaluating a response, an educator may intend to grade only considering important features of a response as specified in the rubric. However, raters may unintentionally consider superficial features like the length of a response or grammatical composition. We fit a hierarchical linear model of ratings of student responses at both the response level and the question level to attempt to explain ratings with superficial features. At the response level, we explore features such as the length of the response, a grammar score, and word frequencies. At the question level, we look at the average number of words across responses and the construct to which the response is related.

The research questions we hope to answer are 1) can we explain expert ratings of constructed responses with non-content related features, and 2) is the effect of certain features different across different items? Limitations of our methods are that we are not able to assume a causal link between these features and their ratings. We cannot make any claim that the expert scorers are in fact using response length when making their judgements. Also, it may be that in some grading scenarios, features such as a grammar score are intended to be considered although with the specific data set we use for this project, this is not the case.

## Methods

### Data set
The data used for this project was from an assessment developed at the Berkeley Evaluation and Assessment Research (BEAR) center consisting of mathematics problem solving items (Arneson et al., 2019). The assessment was created under the BEAR Assessment System (BAS) which outlines principles for assessment development under a construct modelling approach (Wilson, 2023). Construct Modelling describes an iterative process including the creation of a construct map which embodies the theoretical model of student's cognition like understanding of an academic topic, and defining of an outcome space which consists of how responses are mapped to the construct. Student responses were graded by multiple subject-matter experts, and the consistency of ratings was evaluated with an inter-rater reliability score. The data consists of 7,065 student responses from 558 distinct students to 31 items, with the mean number of responses per question being 179. The items relate to four constructs about understanding of algebra: Position, Rate and Acceleration (PRA), Multiple Mathematical Representations (MMR), Modeling Applied Problems (MAP), and Interpreting Mathematical Results (IMR).

### Variables

The variables that we investigate include:
1) The response variable, **rating**. Questions relating to two of the constructs are rated from 0 to 3, and questions relating to the other two constructs are rated from 0 to 4, so 'rating' ranges from 0 to 4.
2) Number of tokens (words) in the response, **num_words.** Tokens are defined as a collection of letters/numbers separated by a space. This variable is a count of the number of tokens in the response.
3) Grammar score of the response, **grammar_score**. This was calculated by a python package, 'LanguageTool', that detects grammar and spelling mistakes including case (first word of a sentence not being capitalized, etc.) using the wrong version of a word such as "too" versus "to", repeating a word twice in a row, using the wrong form of a verb ("ran" instead of "run"), using redundant phrasing, etc.
4) Metric of common word frequency, **word_freq_d.** This is a binary variable indicating whether the response includes any of the five most frequent trigrams - three consecutive words - from the highest-level responses for a question. The dummy variable will be 1 in the case that the response does include at least one of the most frequent phrases, and 0 in the case that the response does not.
5) Average number of tokens per question, **avg_num_words.** This represents the number of tokens, as defined above, averaged across the question level.
6) The construct related to the question, **construct,** will be three dummy variables at the question level with PRA as the reference category.

## Models

In this section, we describe the three models we fit. Interpretations for the model components are as follows: $\beta_1 + \zeta_{1j}$ is the intercept for question j. $\Psi_{11}$ is the variance of the $\zeta_{1j}$s (the intercepts between questions). $\Psi_{22}$ is the variance of $\zeta_{2j}$s (slopes between questions). $\zeta_{1j}$ is the deviation of question j's intercept from the mean intercept, $\beta_1$. $\zeta_{2j}$ is the difference in slope for question j from the overall mean slope of num_words, $\beta_2$. $\varepsilon_{ij}$ is the deviation of unit ij's total residual from the cluster mean. $\beta_1$ represents the mean intercept. $\beta_2$ represents the mean change in rating for a one unit change in num_words. $\beta_2 + \zeta_{2j}$ is the slope of num_words for question j. $\beta_3$ represents the mean change in rating for a one unit change in grammar_score. $\beta_4$ represents the mean difference in rating between word_freq_d being 1 or 0. $\beta_5$ represents the mean change in rating for a one unit change in avg_num_words. $\beta_6$ represents the mean difference in rating between the construct MMR and PRA. $\beta_7$ represents the mean difference in rating between the construct IMR and PRA. $\beta_8$ represents the mean difference in rating between the construct MAT and PRA. $\beta_9$ represents the mean difference in the effect of grammar_score on rating between responses, where a frequently occurring phrase is found (word_freq_d == 1) versus not found (word_freq_d == 0).

<u>Model 1: Random intercepts model with no interaction</u>
First, we fit a random intercepts model where characteristics of the responses include num_words, grammar_score, and word_freq_d, and characteristics of the questions include avg_num_words and construct.

**rating**$_{ij}$ = $\beta_1$ + $\beta_2$**num_words**$_{ij}$ + $\beta_3$**grammar_score**$_{ij}$ + $\beta_4$**word_freq_d**$_{ij}$ + $\beta_5$**avg_num_words**$_j$ + $\beta_6$**construct_MMR**$_j$ + $\beta_7$**construct_IMR**$_j$ + $\beta_8$**construct_MAP**$_j$ + $\zeta_j$ + $\varepsilon_{ij}$

<u>Model 2: Random intercepts model with level 1 interaction</u>
Next, we introduce a response-level interaction between word_freq_d and grammar_score because it might be that the effect of a response's grammar on the rating is different for responses that include frequently occurring phrases in the highest scoring category.

**rating**$_{ij}$ = $\beta_1$ + $\beta_2$**num_words**$_{ij}$ + $\beta_3$**grammar_score**$_{ij}$ + $\beta_4$**word_freq_d**$_{ij}$ + $\beta_5$**avg_num_words**$_j$ + $\beta_6$**construct_MMR**$_j$ + $\beta_7$**construct_IMR**$_j$ + $\beta_8$**construct_MAP**$_j$ + $\beta_9$**grammar_score*word_freq_d**$_{ij}$ + $\zeta_j$ + $\varepsilon_{ij}$

For models 1 and 2, assumptions include that $\zeta_j$ and $\varepsilon_{ij}$ are independent over clusters, j, $\varepsilon_{ij}$ is independent over units, i, $\zeta_j$ and $\varepsilon_{ij}$ are independent of the $x_{ij}$'s and each other, and $\zeta_j$ and $\varepsilon_{ij}$ have zero means and constant variances. Further, the random intercept, $\zeta_j$, can be interpreted as the unobserved heterogeneity between questions. We assume that there is some amount of dependence between responses for the same question. We use restricted maximum likelihood estimation due to small group sizes as the maximum likelihood estimates may underestimate the bias of $\Psi_{11}$, the variance of the random intercept, and the standard error of the estimated values of the $\beta$s.

<u>Model 3: Random coefficients model</u>

Finally, we allow the effect of num_words to vary between questions. If the interaction between word_freq_d and grammar_score is significant in Model 2, we will include this same-level interaction term in the random coefficients model. In addition, we allow the random intercept to be correlated with the random slope.

**rating**$_{ij}$ = $\beta_1$+($\beta_2$ + $\zeta_{2j}$)**num_words**$_{ij}$ +$\beta_3$**grammar_score**$_{ij}$ +$\beta_4$**word_freq_d**$_{ij}$ +$\beta_5$**avg_num_words**$_j$ + $\beta_6$**construct_MMR**$_j$ +$\beta_7$**construct_IMR**$_j$ +$\beta_8$**construct_MAP**$_j$ +$\beta_9$**grammar_score*word_freq_d**$_{ij}$ + $\zeta_{1j}$ +$\varepsilon_{ij}$

Where $\zeta_{1j} \sim N(0, \Psi_{11})$, $\zeta_{2j} \sim N(0, \Psi_{22})$, and cov($\zeta_{1j}$, $\zeta_{2j}$) = $\Psi_{21}$. The added random term, $\zeta_{2j}$, allows for the effect of num_words to vary over questions. The random intercept, $\zeta_{1j}$, is still included in the model as well because even if slopes are allowed to vary, we believe that the intercept will still vary between questions as well.

## Results
In this section, results of all statistical hypothesis tests are interpreted at a significance level of 0.05.

### Model 1 results
In Table 1, we see that the level 1 covariates num_words, grammar_score and word_freq are all significant, but the construct variables (construct_IMR, construct_MAP, and construct_MMR) as well as the level 2 covariate, avg_num_words, are not. Results of the Wald Chi-Square test (statistic=795.84, p<0.001) gives evidence that at least one of the regression coefficients is not equal to zero. In addition, the likelihood ratio test comparing the random intercept model to a one-level ordinary linear regression model (statistic=11.86.67, p<0.001) indicates that including a random coefficient to allow intercepts to vary between questions is an important addition to our model, allowing us to consider the dependency between responses within groups.

**Table 1**
*Fit Results for Model 1: The Random Intercepts Model with No Interaction*

| covariate | coefficient | std. error | Z | P > |Z| |
|---|---|---|---|---|
| num_words | 0.004 | 0.000 | 18.16 | 0.000 |
| grammar_score | -0.089 | 0.020 | -4.43 | 0.000 |
| word_freq | 0.707 | 0.040 | 17.56 | 0.000 |
| construt_IMR | 0.115 | 0.350 | 0.33 | 0.742 |
| construct_MAP | 0.489 | 0.503 | 0.97 | 0.331 |
| construt_MMR | 0.288 | 0.326 | 0.88 | 0.378 |
| avg_num_words | -0.003 | 0.004 | -0.85 | 0.398 |
| intercept | 1.525 | 0.602 | 2.53 | 0.011 |
| **random effect** | **estimate** | **std. error** | | |
| std. dev(intercept) | 0.552 | 0.078 | | |
| st. dev(residual) | 0.986 | 0.009 | | |

### Model 2 results
We do not include the model fit results for Model 2 as the coefficients, std. errors, Z statistics and p-values are very similar to those from Model 1 for covariates besides the interaction term. However, results show the added interaction term, word_freq*grammar_score, should be included in the next model (statistic=-2.23, p=0.026).

### Model 3 results
In Table 3, the chi-squared likelihood ratio test for comparing Model 2 to Model 3 shows that the random coefficient model is significantly different from the random intercept model (statistic=213.14, p<0.001). The estimated mean change in rating for adding an additional word to the response is 0.010 (p<0.001), indicating that longer responses tend to be rated higher. In addition, for a one unit increase in grammar_score, corresponding to more grammar mistakes, and when word_freq is 0 (phrases frequently found in high scoring responses are not found in the response), the estimated mean change in score is -0.084 (p<0.001). So, more grammar mistakes tend to correspond to a lower rating. The estimated mean difference in rating for responses that include the top phrases versus responses that do not include the top phrases, when grammar_score is 0, is 0.706 (p<0.001), so the inclusion of certain phrases in a response corresponds to a higher rating. The significance of the interaction between word_freq and grammar_score (p=0.018) tells us that the effect of grammar in a student's response is different for when certain key phrases exist in the response, versus when they do not.

The level 2 covariates and the construct variables are not significant (for all, p>0.05) indicating that the average number of words at the question level and the construct to which the question belongs are not useful features of a student's response to explain rating. Further, the correlation between the random effects is -0.761, signifying that there is a tendency for clusters with larger intercepts to have smaller slopes. In addition, the st. dev of the level 1 residual (0.956) represents the amount of scatter around the question-specific regression lines.

**Table 3**
*Fit Results for Model 3: The Random Coefficients Model with an Interaction*

| covariate | coefficient | std. error | Z | P > |Z| |
|---|---|---|---|---|
| num_words | 0.010 | 0.002 | 4.66 | 0.000 |
| grammar_score | -0.084 | 0.021 | -4.00 | 0.000 |
| word_freq | 0.706 | 0.043 | 16.51 | 0.000 |
| word_freq*grammar_score | -0.119 | 0.050 | -2.36 | 0.018 |
| construt_IMR | 0.049 | 0.337 | -0.15 | 0.884 |
| construt_MAP | 0.712 | 0.485 | 1.47 | 0.142 |
| construt_MMR | 0.028 | 0.314 | 0.09 | 0.929 |
| avg_num_words | -0.004 | 0.004 | -1.03 | 0.305 |
| intercept | 1.448 | 0.591 | 2.45 | 0.014 |
| **random effect** | **estimate** | **std. error** | | |
| std. dev(num_words) | 0.012 | 0.002 | | |
| st. dev(residual) | 0.812 | 0.171 | | |
| corr(num_words, intercept) | -0.761 | 0.119 | | |
| st. dev(residual) | 0.956 | 0.009 | | |

## Discussion

The results confirm our original hypothesis - that features unrelated to the actual content of a student's response like the number of words in a response, and grammar, play a role in explaining the final rating of a student's response. In addition, we saw from the random intercept model that allowing for separate regression intercepts for different questions was useful. Finally, for one of the features, the length or number of words in a response, the effect on rating was different for different questions. Implications of these findings are important to educators who consider the ratings of constructed response questions to evaluate students, especially in high-stakes scenarios. Although professionals who are rating open-ended responses take great care to provide unbiased, consistent ratings, it might be that certain factors, unrelated to what is intended to be rated, are being evaluated.

Our results are not causal, but they may lead us to speculate about fairness issues related to grading short response items. The effect of grammar score may correlate with an effect for non-native English speakers, as grammar can come as a challenge for English as a Second Language (ESL) learners. ESL students who have a high understanding of the subject-matter content may be penalized for grammar mistakes unintentionally. Further, simply including more text in a response or including key phrases without connecting ideas do not necessarily indicate a higher level of understanding, although human raters may hold these unconscious biases when grading responses. We recognize that grading short response items can be an arduous task and educators often lack the time and resources necessary to rate carefully. For this project, we intentionally used data with meticulously rated responses by multiple raters to illuminate that these effects are present even under strict controls. As education professionals, we should be aware that such effects are present, and critical of our own grading practices.